\documentclass{article}

\usepackage[english]{babel}

\usepackage[letterpaper,top=2cm,bottom=2cm,left=3cm,right=3cm,marginparwidth=1.75cm]{geometry}

\usepackage{amsmath}
\usepackage{amssymb}
\usepackage{graphicx}
\usepackage[colorlinks=true, allcolors=blue]{hyperref}
\usepackage{authblk}

\usepackage{bm}
\newcommand{\dd}{\mathrm{d}}
\newcommand{\newhl}{\bm{h}_l}
\newcommand{\hll}{\bm{h}_{l-1}}
\newcommand{\wHl}{\widetilde{H}_l}
\newcommand{\wHll}{\widetilde{H}_{l-1}}
\newcommand{\mm}{\bm{m}}
\newcommand{\wh}{k}
\newcommand{\wq}{\widetilde{q}}
\newcommand{\wtp}{\widetilde{p}}
\newcommand{\wKl}{\widetilde{K}_l}
\newcommand{\wKll}{\widetilde{K}_{l-1}}
\newcommand{\whl}{\bm{\wh}_l}

\newcommand{\opt}{\mathrm{opt}}

\title{Performance Bounds of Ranging Precision in SPAD-Based dToF LiDAR}
\author{
Hao Wu $^{1,2,}$\thanks{wuhao\_zju@zju.edu.cn}, Shiyi Sun$^{1}$, Lijie Zhao$^{1}$, Yingyu Wang$^{1}$, Limin Tong $^{2}$ and Linjie Shen $^{1}$}
\affil{$^{1}$ \quad Hikvision Research Institute, Hangzhou 310051, China;}
\affil{$^{2}$ \quad State Key Laboratory of Modern Optical Instrumentation, College of Optical Science and Engineering, Zhejiang University, Hangzhou 310027, China;}

\begin{document}
\date{}
\maketitle

\begin{abstract}
Lidar with direct time-of-flight (dToF) technology based on single-photon avalanche diode detectors (SPAD) has been widely adopted in various applications.
However, a comprehensive theoretical understanding of its fundamental ranging performance limits—particularly the degradation caused by pile-up effects due to system dead time and the potential benefits of photon-number-resolving architectures—remains incomplete.
In this work, the Cramér-Rao lower bound (CRLB) for dToF systems is theoretically derived accounting for dead time effects, generalized to SPAD detectors with photon-number-resolving capabilities, and are further validated through Monte Carlo simulations and maximum likelihood estimation.
Our results reveal that pile-up not only reduces the information contained within individual ToF but also introduces statistical coupling between distance and photon flux rate, further degrading ranging precision.
The derived CRLB is used to determine the optimal optical photon flux, laser pulse width, and ToF quantization resolution that yield the best achievable ranging precision.
The analysis further quantifies the limited performance improvement enabled by increased photon-number resolution, which exhibits rapidly diminishing returns.
These findings provide theoretical guidance for the design of dToF systems and the selection of their optimal operating points.
\end{abstract}

\section{Introduction}

Laser detection and ranging (LiDAR) technology has gained considerable attention in a wide range of applications, including autonomous driving \cite{liangEvolutionLaserTechnology2024,kumagai73189x600BackIlluminated2021,rappAdvancesSinglePhotonLidar2020}, robotics \cite{khanComparativeSurveyLiDARSLAM2021}, and remote sensing \cite{diabDeepLearningLiDAR2022,wangChallengesOpportunitiesLidar2021}.
At its core, LiDAR determines object distances by measuring the time-of-flight (ToF) of laser pulses reflected from targets.
Among the various ToF methods, direct ToF (dToF) approaches—by directly measuring the travel time of laser pulses—are particularly attractive due to their structural simplicity and rapid measurement capabilities \cite{villaSPADsSiPMsArrays2021,maReviewToFbasedLiDAR2024}.

A key component of recent advances in dToF systems is the single-photon avalanche diode (SPAD).
Operating in Geiger mode, SPADs are capable of detecting individual photons with high temporal resolution and low noise, even under photon-starved conditions \cite{wang641283DStacked2024,morimoto32Megapixel3DStacked2021}.
However, each SPAD pixel undergoes a so-called quenching process after being triggered, during which it recovers from avalanche and prepares for the next detection event \cite{zappaSPICEModelingSingle2009}.
This recovery period, known as the dead time, represents an interval during which the pixel is inactive and incapable of registering new photon events.
Moreover, a single SPAD pixel is inherently binary in its response—it only signals the presence or absence of a photon, without quantifying the number of incident photons.
To circumvent this limitation, recent efforts have introduced macro-pixel architectures composed of multiple subpixels \cite{zhang2401603DStacked2022,padmanabhan74256x1283DStacked2021,kumagai73189x600BackIlluminated2021,hutchingsReconfigurable3DStackedSPAD2019}.
In such configurations, a macro-pixel consisting of $s$ subpixels can potentially register between 0 and $s$ triggered subpixels in a single measurement, thereby endowing the detector with $s$-photon-number resolution ($s$-PNR).
This additional degree of photon number information not only extends the system's dynamic range but also holds promise for improving ranging precision by capturing richer photon statistics.

Another critical component of dToF systems is the time-to-digital converter (TDC), which quantizes the temporal interval between the laser emission and the SPAD detection event \cite{incoronatoStatisticalModellingSPADs2021a,xiaReviewAdvancementsTrends2024}.
A key figure of merit for TDCs is their time resolution, which governs the precision of ToF quantization.
However, TDCs are also subject to a form of dead time, imposed by circuit constraints, during which further conversions are temporarily inhibited.
Early TDCs were only capable of registering a single event per measurement cycle, which can be considered as an infinite dead time.
Advances in circuit design have since allowed for multiple ToF conversions within a single cycle \cite{wang79PsResolution2024,hutchingsReconfigurable3DStackedSPAD2019}.
The so-called multi-event TDC increases the available photon timing information and, in principle, can further enhance the system's ranging accuracy.

In practice, photon detections are inevitably contaminated by noise, including ambient irridiation and circuit noise.
To improve the signal-to-noise ratio, modern dToF systems typically emit a sequence of laser pulses and compile a histogram of the corresponding quantized ToF values in which the bin width equals the TDC resolution.
Ideally, the bin with the highest count corresponds to the estimated ToF.
To further improve estimation accuracy to sub-bin resolution, previous studies have proposed various techniques such as curve-fitting to achieve sub-bin precision by fully exploiting the histogram peak shape \cite{gyongyHighspeed3DSensing2020,koernerModelsDirectTimeofFlight2021}.

A rigorous theoretical analysis of ranging precision is essential for guiding the architectural and algorithmic design of dToF LiDAR systems.
While prior work has provided valuable insights into this area \cite{scholesFundamentalLimitsDepth2023,koernerModelsDirectTimeofFlight2021,scholesRobustFrameworkModelling2024}, two key aspects remain insufficiently explored.
First, the distortion of histogram peak shapes induced by dead time is not fully accounted for, which leads to a deviation between the location of the histogram peak and the ToF corresponding to the true distance, a phenomenon commonly referred to as the "walk error".
Several calibration methods have been proposed to correct walk errors \cite{rappDeadTimeCompensation2019,choiWalkErrorCompensation2022,huangkeTheoreticalModelCorrection2018,yangMethodRangeWalk2022}, yet the fundamental limits of such correction remain unknown.
Second, the benefits conferred by macro-pixel architectures with photon-number resolution have not been rigorously characterized from an theoretical perspective.

In this work, the ranging performance of dToF systems is theoretically analyzed based on the fisher information theory by explicitly incorporating the impact of dead time, and the result is generalized to include photon-number-resolving SPAD detectors.
Specifically, we derive the probability density functions of detection histograms under various dToF configurations and compute the corresponding Cramér-Rao lower bound (CRLB) on ranging precision.
To validate the theoretical results, Monte Carlo simulations are conducted to generate detection histograms, from which maximum likelihood estimation is performed to infer distance.
The simulation results exhibit excellent agreement with theory, confirming the robustness and accuracy of our modeling approach.

Our derivations reveal that the presence of dead time leads to additional degradation in the theoretical ranging precision compared to idealized models.
Furthermore, the analysis reveals that excessive received pulse intensities can severely deteriorate ranging performance, whereas an optimal received intensity exists that minimizes the CRLB under given system constraints.
Nevertheless, an optimal precision of approximately $0.53\tau/\sqrt{N}$ for a 1-PNR detector remains theoretically achievable by appropriately selecting the laser pulse width and the received photon flux rate, where $\tau$ is TDC resolution and $N$ is the number of laser pulses.
In the case of $s$-PNR SPAD macro-pixels, it is found that the full potential of photon-number resolution can be realized only when the total number of TDC triggers per measurement cycle is also available in addition to the commonly used histogram of total sub-pixel counts.
Moreover, increasing photon-number resolution yields only marginal improvements in the system's theoretical best-case ranging performance.

In conclusion, the presented results clarify the theoretical performance boundaries of SPAD-based dToF LiDAR and offer important guidelines for the design of future high-precision, photon-efficient ranging systems.

\section{Materials and Methods}
\subsection{Detection Model}
We begin by modeling SPAD-based dToF LiDAR systems without incorporating the technical details of acquisition electronics or specific measurement techniques.
While prior work has derived key results such as the expected histogram counts via single-pulse detection analysis \cite{incoronatoStatisticalModellingSPADs2021,padmanabhanModelingAnalysisDirect2019a}, here the results are derived from first principles by obtaining the histogram's full probability distribution, thereby providing the basis for CRLB derivation in next section.
For simplicity, it is assumed that the SPAD dead time is equal to the TDC dead time and this unified dead time is denoted as $T$.
This assumption avoids the complex histogram behaviors that can arise when the SPAD dead time is shorter than that of the TDC \cite{rappHighfluxSinglephotonLidar2021a}, which would reduce the effective photon count and degrade the ranging performance.
Moreover, since the focus of this work is on analyzing the fundamental limits of ranging accuracy, the model does not account for non-ideal effects such as afterpulsing, crosstalk, or other circuit-level impairments that may degrade ranging performance in practical implementations \cite{wojtkiewiczReviewBackSideIlluminated2024}.

Let the normalized waveform of the emitted laser pulse be denoted by $f(t)$, such that $\operatorname{max}\{f(t)\}=1$ and $f(t<0)=0$.
To prevent a ambiguity in ToF estimation introduced by a single laser pulse from generating multiple peaks in the histogram, it is further assumed that the duration of the laser pulse is shorter than the system's dead time, i.e. $f(t>T-1)=0$.

Upon reflection from a target located at distance $d$, the pulse returns to the dToF system and is focused onto the SPAD pixel, generating a signal photon flux of $R f(t-t0)$, where $t_0=2d/c$ is the round-trip ToF and $R$ denotes the peak photon flux rate absorbed by the SPAD, which is determined by the peak optical power of the emitted pulse, the reflectivity of the target surface, and the detection efficiency of the SPAD, etc. 
The mean photon count of the signal within the time interval of the $i$-th histogram bin is expressed by
\begin{equation}
    S_i = R \int_{i\tau}^{(i+1)\tau} f(t-t_0) \dd t,
\end{equation}
where $\tau$ denotes the bin width of the histogram, i.e., the resolution of the TDC.
Similarly, the average photon count of noise is given by $b = R_n \tau$, where $R_{n}$ denotes the average photon flux of noise absorbed by the SPAD.
For coherent light, the photon counts follow a Poisson distribution.
Therefore, the probability of no detection event occurring in the $i$-th bin, denoted as $p_i$, is given by $p_i = e^{-S_i-b}$.
Correspondingly, the probability of a detection event occurring in the $i$-th bin is given by $q_i = 1 - p_i$.

For a SPAD detector without PNR capability, i.e., a 1-PNR SPAD, let the system perform $N$ independent pulse measurements, accumulating the histogram $\bm{h} = [h(0), h(1), \dots]$.
If the counts in the previous $T$ bins are known as $h_{i-1},h_{i-2},\dots,h_{i-T}$, then due to the dead time constraint, the count $h(i)$ in the $i$-th bin follows a conditional binomial distribution:
\begin{equation}
    \label{eq:conP}
    P\{h(i)=h_i | h(i-1)=h_{i-1}, h(i-2)=h_{i-2}, ..., h(i-T)=h_{i-T}\}\sim \mathcal{B}(N'_i, q_i),
\end{equation}
where $N'_i$ is defined as
\begin{equation}
    \label{eq:Np}
    N'_i=N-\sum_{j=1}^{T} h_{i-j}.
\end{equation}
For simplicity, this conditional probability is denoted as $H_i(h_i | h_{i-1}, h_{i-2}, ..., h_{i-T})$.
With the law of total expectation, the expectation of $P\{h(i)=h_i\}$ is given by

\begin{equation}
\label{eq:Ehi}
E[h(i)] = q_i E[N'_i] = q_i (N - \sum_{j=1}^{T} E[h(i-j)]).
\end{equation}

Defining $E[h(i)] = N Q_i = N q_i F_i$, where $Q_i = q_i F_i$ represents the expected count in the $i$-th bin for a single measurement, it can be seen that this expectation is proportional to the probability of generating a detection event, $q_i$, multiplied by the correction factor $F_i$, which accounts for the reduced number of measurable pulses due to dead time, reflecting the so-called “pile-up” effect.
Compared with Eq.~\ref{eq:Ehi}, the expression for $F_i$ can be derived as
\begin{equation}
    \label{eq:Fdef}
    F_i = 1 - \sum_{j=1}^{T} Q_{i-j},
\end{equation}
which indicates that the $F_i$ equals the probability that the $i$-th bin is not within the dead time caused by triggers in the preceding $T$ bins.
It can be further expressed as a recursive relation
\begin{equation}
    \label{eq:Frec}
    F_i = F_{i-1} - Q_{i-1} + Q_{i-T-1} = F_{i-1}p_{i-1} + Q_{i-T-1}.
\end{equation}
If the system uses a single-TDC architecture, where the TDC triggers only once per laser pulse measurement cycle, then the recursive relation can be written as
\begin{equation}
    \label{eq:F0def}
    F_i = F_{i-1} - Q_{i-1} = F_{i-1}p_{i-1}.
\end{equation}

For the boundary conditions of the above recursion, before each measurement cycle starts, the SPAD is in a steady state with only background noise present, therefore the probability of the SPAD being triggered at each "negative bin" with $i<0$ should be equal.
Denoting $q_b = 1 - e^{-b}$ the probability of a trigger in one bin under background noise without accounting for dead time effect, there are three possible conditions for a single "negative bin":
1) Within the dead time caused by the preceding detection event;
2) Not affected by dead time, and generating a detection event;
3) Not affected by dead time, and not generating a detection event.
Assuming $A$ is the probability of a "negative bin" not affected by dead time, then the probabilities of the above three conditions are:
$A q_b T$ (affected by dead time), $A q_b$ (not affected and triggered), and $A (1 - q_b)$ (not affected and no trigger).
With normalized condition $A q_b + A (1 - q_b) + A q_b T = 1$, it results in $A = 1/(1 + q_b T)$.
Therefore, the joint probability distribution for an arbitrary continuous span of $T+1$ "negative bins" before the measurement cycle starts follows a multinomial distribution:
\begin{eqnarray}
&&P\{h(j)=h_j, h(j+1)=h_{j+1}, ..., h(j+T)=h_{j+T}\}\nonumber\\
&=&\binom{N}{h_j\ h_{j+1}\ \cdots\ h_{j+T}}\left(\frac{q_b}{1+q_b T}\right)^{\sum_{k=j}^{j+T} h_k}\left(\frac{1-q_b}{1+q_b T}\right)^{N-\sum_{k=j}^{j+T} h_k},
\end{eqnarray}
where $\binom{N}{h_j\ h_{j+1}\ \cdots\ h_{j+T}}$ is the multinomial coefficient
\begin{equation}
    \binom{N}{h_j\ h_{j+1}\ \cdots\ h_{j+T}} = \frac{N!}{h_j! h_{j+1}!\cdots h_{j+T}!(N-\sum_{k=0}^{T} h_k)!}.
\end{equation}
And the expected histogram values for the virtual bins is
\begin{eqnarray}
&&E[h(i<0)]\nonumber\\
&=& \sum_{\mm} m_i P\{h(-1)=m_1, h(-2)=m_2, ..., h(-T-1)=m_{T+1}\}\nonumber\\
&=& \left(\frac{1}{1+q_b T}\right)^N \sum_{\mm} m_i \binom{N}{m_0\ \cdots\ m_T} q_b^{m'_{T}} (1-q_b)^{N-m'_{T}}\nonumber\\
&=& \left(\frac{1}{1+q_b T}\right)^N \left.\frac{\partial[q_b T + q_b x + (1-q_b)]^N}{\partial x}\right|_{x=1}\nonumber\\
&=& N \frac{q_b}{1+q_b T}~,
\end{eqnarray}
where $\mm=(m_1, \dots, m_T)$ and $m'_T=\sum_{i=1}^{T+1} m_i$.
With $E[h(i<0)]=N Q_i$, the above result implies that that $Q_{i<0} = q_b/(1 + q_b T)$.
Therefore, we obtain $F_{i\leq T} = 1/(1 + q_b T)$ for both multi- and single-event TDC architectures, which aligns with the aboving $A$ and also results previously reported in the literature \cite{incoronatoStatisticalModellingSPADs2021a}.

For $s$-PNR SPADs, which consist of $s$ identical sub-pixels, it is assumed that the incident light is uniformly distributed among all sub-pixels, yielding an instantaneous photon flux rate per sub-pixel of $R'=R/s$.
During the time interval corresponding to the $i$-th bin, the probability that a sub-pixel does not and does generate a detection event is given by $\wtp_i = e^{-(S_i+b)/s}$ and $\wq_i = 1-\wtp_i$, respectively.
To simplify the analysis and enable analytical derivation, we further assume synchronous triggering across sub-pixels.
For each laser pulse measurement, the detector records the number of sub-pixels triggered and accumulates this count into the histogram, denoting as $\wh(i)$.

Similar to Eq.~\ref{eq:conP}, due to the constraint imposed by dead time, calculating the total count $\wh(i)$ in the $i$-th bin requires knowledge of the number of TDC triggers that occurred within the preceding $T$ bins.
These TDC trigger counts are equivalent to the counts obtained by treating the s-PNR SPAD as a conventional 1-PNR SPAD, so they are also denoted as $h_i$.
The conditional probability of observing a total of $\wh_i$ triggered subpixels in the $i$-th bin, given $h(i)=h_i$ TDC triggers, is expressed as:

    \begin{equation}
    \label{eq:Ps}
    P\{\wh(i)=\wh_i | h(i)=h_i, h(i-1)=h_{i-1}, ..., h(i-T)=h_{i-T}\} = \binom{N'_i}{h_i} J_s(h_i, \wh_i) \wq_i^{\wh_i} \wtp_i^{s N'_i - \wh_i},
\end{equation}

where $\binom{N'_i}{h_i}$ is the binomial coefficient, $N'_i$ is defined as in Eq.~\ref{eq:Np}, and $J_s(h_i, \wh_i)$ denotes the number of ways to allocate $\wh_i$ triggered subpixels across $h_i$ TDC events, under the constraint that no more than $s$ subpixels can be triggered per event, satisfying
\begin{equation}
    \sum_{h_i=0}^{N'_i} \binom{N'_i}{h_i} J_s(h_i, \wh_i) = \binom{s N'_i}{\wh_i}.
\end{equation}
Therefore, marginally, the total photon count $\wh(i)$ follows the binomial distribution:
\begin{equation}
    P(\wh(i) | N'_i) \sim \mathcal{B}(s N'_i, \wq_i).
\end{equation}
Using the law of total expectation and referring to Eq.~\ref{eq:Ehi}, the mean of $\wh(i)$ across $N$ measurements can be written as
\begin{equation}
    \label{eq:Ewhi}
    E[\wh(i)] = s N Q_i = s N F_i \wq_i,
\end{equation}
where $F_i$ remains defined as in Eq.~\ref{eq:Fdef}.

\begin{figure}[!h]
\centering % Only used for preprints
\includegraphics[width=7 cm]{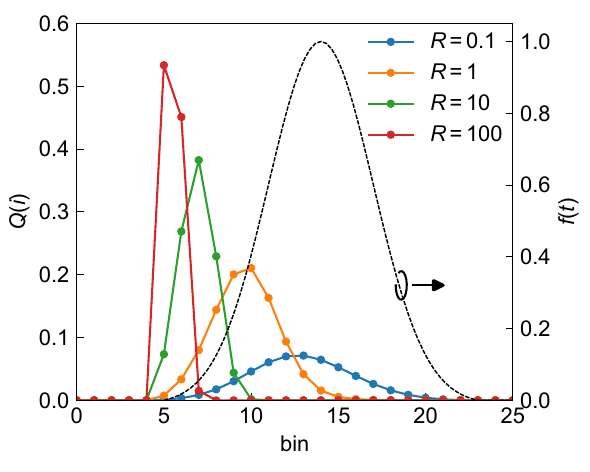}
\caption{Theoretical histogram expectation and pile-up effect.
Black dashed line: the theoretical laser pulse waveform $f(t-t_0)$ corresponding to a ToF of $t_0=5$.
Solid lines: expected single-pulse histograms $Q_i$ under different peak received photon flux rates with $R=0.1$ (blue), 1 (orange), 10 (green), and 100 (red).\label{fig:walk-error}}
\end{figure}

Figure~\ref{fig:walk-error} illustrates the expected histograms $Q_i$ corresponding to a Gaussian laser pulse reflected from targets located at the same distance $t_0$, but with varying reflectivity levels, resulting in different photon flux rates $R$, in a dToF system using a 1-PNR SPAD.
As $R$ increases, the temporal bias of the detection histogram progressively to the left and becomes narrower, indicating a more severe pile-up effect caused by the detector's dead time.
If the target distance is estimated directly from the peak position in the histogram, this bias leads to a walk error, where targets at the same physical distance but with different reflectivities are erroneously estimated to be at different distances.

\unskip

\subsection{Cramér-Rao lower bound}
Based on the system's probabilistic model derived earlier, the theoretical ranging performance—specifically, the Cramér-Rao lower bound (CRLB) for the variance of the ToF estimate $\operatorname{Var}\{\widehat{t_0}\}$—can be calculated.
To evaluate the fundamental performance limit of detection, it is essential to minimize the number of unknown system parameters other than $t_0$.
As discussed in the previous section, the system output (i.e., the histogram) is primarily influenced by two categories of parameters.
The first category consists of parameters that are intrinsic to the detection system and independent of the measurement environment, including the dead time $T$, TDC resolution $\tau$, and the laser pulse waveform $f(t)$, which all can be accurately characterized through system design and calibration.
The second category comprises environment-dependent parameters, such as the background noise level $R_n$, the target distance (i.e., ToF $t_0$), and the peak received photon rate $R$.
Among these parameters, the background noise rate $R_n$, which includes ambient light and dark noise, can be reliably estimated via long-duration passive measurements by the dToF system (e.g., acquiring histograms without emitting laser pulses), and thus is assumed to be known.
Given the lack of prior knowledge about the target object, the set of unknown parameters to be estimated is defined as $\bm{\theta} = (t_0, R)$.

Assuming that the histogram contains only a single target-induced peak located at $t_0$, the probability of observing a specific histogram realization with $h(0) = h_0, h(1) = h_1, ..., h(l)=h_l$ is given by

\begin{eqnarray}
\label{eq:P1}
&&P(\newhl;\bm{\theta})\nonumber\\
&=&P\{h(0) = h_0, h(1) = h_1, ..., h(l) = h_l\} \nonumber \\
&=&\sum_{\bm{m}} E[P\{h(0) = h_0, h(1) = h_1, ..., h(l) = h_l | h(-1)=m_1,h(-2)=m_2,...,h(-T)=m_T\} \times \nonumber\\
&&P\{h(-1)=m_1,h(-2)=m_2,...,h(-T)=m_T, h(-T-1)=m_{T+1}\}] \nonumber \\
% &=& E_m[P\{h(0) = h_0 | h(-1)=m_1,h(-2)=m_2,...h(-T)=m_T\} \nonumber \\
% &&\times P\{h(1) = h_1 | h(0)=h_0, h(-1)=m_1,h(-2)=m_2,...h(-T+1)=m_{T-1}\}\cdots] \nonumber \\
&=& E_{\mm}[H_0 H_1 \cdots H_l]\nonumber\\
&\overset{\mathrm{def}}{=}& \wHl,
\end{eqnarray}

where $\newhl = (h_0, h_1, ..., h_l)$ and $E_{\mm}[\cdot]$ is the expection over all possible $\mm$.
The Fisher information matrix is defined as \cite{kayFundamentalsStatisticalSignal1993}
\begin{equation}
I_{ij}(\bm{\theta}) = \sum_{\newhl}P(\newhl;\bm{\theta})\left(\frac{\partial\ln{P(\newhl;\bm{\theta})}}{\partial \theta_i}\right)\left(\frac{\partial\ln{P(\newhl;\bm{\theta})}}{\partial \theta_j}\right).
\end{equation}
The simplification for an arbitrary element in the matrix can be expressed as

\begin{eqnarray}
\label{eq:Psim}
&&\sum_{\newhl} P(\newhl;\bm{\theta})\left(\frac{\partial\ln{P(\newhl;\bm{\theta})}}{\partial \theta_i}\right)\left(\frac{\partial\ln{P(\newhl;\bm{\theta})}}{\partial \theta_j}\right) \nonumber \\
% & = & \sum_{\newhl}\frac{1}{P(\newhl;\bm{\theta})}\left(\frac{\partial P(\newhl;\bm{\theta})}{\partial \theta_i}\right)\left(\frac{\partial P(\newhl;\bm{\theta})}{\partial \theta_j}\right) \nonumber \\
& = & \sum_{\newhl}\frac{1}{\wHl} \left(\frac{\partial \wHl}{\partial \theta_i}\right)\left(\frac{\partial \wHl}{\partial \theta_j}\right) \nonumber \\
& = & \sum_{\newhl}\frac{1}{\wHl} \left(\wHll\frac{\partial H_l}{\partial \theta_i} + \frac{\partial \wHll}{\partial \theta_i}H_l\right)\left(\wHll\frac{\partial H_l}{\partial \theta_j} + \frac{\partial \wHll}{\partial \theta_j}H_l\right) \nonumber \\
& = & \sum_{\newhl} \left(\wHll\frac{1}{H_l}\frac{\partial H_l}{\partial \theta_i} \frac{\partial H_l}{\partial \theta_j} + \frac{1}{\wHll} H_l \frac{\partial \wHll}{\partial \theta_i} \frac{\partial \wHll}{\partial \theta_j} + \left(\frac{\partial \wHll}{\partial \theta_i}\frac{\partial H_l}{\partial \theta_j} + \frac{\partial H_l}{\partial \theta_i}\frac{\partial \wHll}{\partial \theta_j}\right)\right).
\end{eqnarray}

For the first term, the derivation term can be simplified according to Eq.~\ref{eq:conP} as
\begin{equation}
    \label{eq:I1dev}
\frac{\partial H_l}{\partial \theta_i} = H_l \frac{h_l - N'_l q_l}{p_l q_l} \frac{\partial q_l}{\partial \theta_i}.
\end{equation}
So the first term results in [see Appendix A]
\begin{equation}
    \label{eq:I1term}
\sum_{\newhl}\wHll\frac{1}{H_l}\frac{\partial H_l}{\partial \theta_i} \frac{\partial H_l}{\partial \theta_j} \nonumber = N \frac{F_l}{p_l q_l} \frac{\partial q_l}{\partial \theta_i}\frac{\partial q_l}{\partial \theta_j}.
\end{equation}
The second term simplifies to the sum of the Fisher information from bin 0 to bin $l-1$: 
\begin{eqnarray}
    \label{eq:I2term}
&&\sum_{\newhl}\frac{1}{\wHll} H_l \frac{\partial \wHll}{\partial \theta_i} \frac{\partial \wHll}{\partial \theta_j} \nonumber \\
&=& \sum_{\hll}\frac{1}{\wHll} \frac{\partial \wHll}{\partial \theta_i} \frac{\partial \wHll}{\partial \theta_j} \sum_{h_l}H_l \nonumber \\
&=& \sum_{\hll}\frac{1}{\wHll} \frac{\partial \wHll}{\partial \theta_i} \frac{\partial \wHll}{\partial \theta_j} \nonumber \\
&=& \sum_{\hll} P(\hll;\bm{\theta})\left(\frac{\partial\ln{P(\hll;\bm{\theta})}}{\partial \theta_i}\right)\left(\frac{\partial\ln{P(\hll;\bm{\theta})}}{\partial \theta_j}\right).
\end{eqnarray}
For the third term is simply zero [see Appendix A].

Combining the above results, the Fisher information simplifies to
\begin{equation}
    \label{eq:IijFl}
    I_{ij}(\bm{\theta}) = N \sum_{k=0}^l \frac{F_k}{p_k q_k} \frac{\partial q_k}{\partial \theta_i}\frac{\partial q_k}{\partial \theta_j}.
\end{equation}
Let $l_1 = \lfloor t_0/\tau \rfloor$ denote the first bin containing the returning laser pulse and $l_2 = l_1 + T$.
Since $\partial q_i/\partial t_0\neq 0$ only when $l_1 \leq i \leq l_2$,
the Fisher information matrix can be expressed as
\begin{equation}
    \label{eq:I1}
    I_{ij}(\bm{\theta}) = N \mathcal{I}_{ij}(\bm{\theta}) = N \sum_{k=l_1}^{l_2} \frac{F_k}{p_k q_k} \frac{\partial q_k}{\partial \theta_i}\frac{\partial q_k}{\partial \theta_j},
\end{equation}
where $\bm{\mathcal{I}}$ denotes the Fisher information corresponding to a single pulse measurement.
It can be noted that each term in the sum exactly corresponds to the Fisher information contributed by the $i$-th histogram bin when the counts follow a binomial distribution $\mathcal{B}(N F_i(\bm{\theta)}, q_i(\bm{\theta)})$ with respect to estimating the parameter $\bm{\theta}$.

The variance of the estimator for the ToF $t_0$ is given by the $(1, 1)$-th element of the inverse Fisher information matrix:
\begin{equation}
    \operatorname{Var}\{\widehat{t_0}\} = \left[\bm{I}(t_0, R)^{-1}\right]_{11} = \frac{1}{N\left[\mathcal{I}_{11} - \dfrac{\mathcal{I}_{12}^2}{\mathcal{I}_{22}}\right]}
\end{equation}
It is clear that the variance of the estimator for $t_0$ decreases inversely with $N$.
Define the correlation coefficient between the two parameters as
\begin{equation}
    \rho^2 = \frac{\mathcal{I}_{12}^2}{\mathcal{I}_{11} \mathcal{I}_{22}},
\end{equation}
then the above expression can be rewritten as
\begin{equation}
    \label{eq:vart0}
    \operatorname{Var}\{\widehat{t_0}\} = \frac{1}{N \mathcal{I}_{11}(1-\rho^2)}.
\end{equation}

From the Eq.~\ref{eq:I1} and \ref{eq:vart0}, the presence of dead time degrades the variance of $t_0$ estimation in two ways.
On the one hand, each histogram bin's individual Fisher information contribution is weighted by the factor $F_i$.
Since $F_i \leq 1$ and is monotonically decreasing, bins corresponding to later times contribute less information.
On the other hand, the dead time induced pile-up effect introduces correlation $\rho$ between the estimates of $t_0$ and $R$, which further degrades the variance of $t_0$ estimation as $\rho$ increases.
When $\rho=0$, the estimations of the two parameters are statistically independent, yielding $\operatorname{Var}\{\widehat{t_0}\}=1/(N \mathcal{I}_{11})$, meaning the variance of $t_0$ is unaffected by whether $R$ is known or not.
As $\rho$ increases, the correlation between the two parameter estimates increases, which leads to degraded estimation performance for $t_0$.
When $\rho$ is approaching 1, the estimates become fully correlated and $\operatorname{Var}\{\widehat{t_0}\}$ diverges implying that effective estimation of $t_0$ is impossible.

Similarly, for the $s$-PNR SPAD detector, we denote Eq.~\ref{eq:Ps} as $K_i(\wh_i|h_i, h_{i-1}, ..., h_{i-T})$ and the probability of observing a specific histogram with photon number counts $\bm{\wh}=[\wh(1) = \wh_1, \wh(2) = \wh_2, ...]$ and corresponding TDC trigger counts $\bm{h}=(h(1) = h_1, h(2) = h_2, ...)$ is given by
\begin{eqnarray}
\label{eq:P2}
&&P(\whl, \newhl;\bm{\theta}) \nonumber \\
&=&P\{\wh(1) = \wh_1, h(1) = h_1, \wh(2) = \wh_2, h(2) = h_2, ...\} \nonumber \\
&=& E_m[K_1 K_2 \cdots K_l] \nonumber \\
&\overset{\mathrm{def}}{=}& \wKl.
\end{eqnarray}
And the fisher information matrix is defined as
\begin{equation}
    I_{ij}(\bm{\theta}) = \sum\limits_{\whl, \newhl}P(\whl, \newhl;\bm{\theta})\left(\frac{\partial\ln{P(\whl, \newhl;\bm{\theta})}}{\partial \theta_i}\right)\left(\frac{\partial\ln{P(\whl, \newhl;\bm{\theta})}}{\partial \theta_j}\right).
\end{equation}

Similarly to Eq.~\ref{eq:Psim}, each element of the Fisher information matrix can be decomposed into three terms as follows:

\begin{eqnarray}
    &&\sum\limits_{\whl, \newhl}P(\whl, \newhl;\bm{\theta})\left(\frac{\partial\ln{P(\whl, \newhl;\bm{\theta})}}{\partial t_0}\right)\left(\frac{\partial\ln{P(\whl, \newhl;\bm{\theta})}}{\partial R}\right) \nonumber \\
    & = & \sum_{\whl, \newhl} \left(\wKll\frac{1}{K_l}\frac{\partial K_l}{\partial \theta_i} \frac{\partial K_l}{\partial \theta_j} + \frac{1}{\wKll} K_l \frac{\partial \wKll}{\partial \theta_i} \frac{\partial \wKll}{\partial \theta_j} + \left(\frac{\partial \wKll}{\partial \theta_i}\frac{\partial K_l}{\partial \theta_j} + \frac{\partial K_l}{\partial \theta_i}\frac{\partial \wKll}{\partial \theta_j}\right)\right)
\end{eqnarray}

The derivative term in the first term is calculated as
\begin{equation}
    \label{eq:K1}
    \frac{\partial K_l}{\partial \theta_i} = K_l \frac{\wh_l - s N'_l \wq_l}{\wtp_l \wq_l} \frac{\partial \wq_l}{\partial \theta_i}.
% \frac{\partial K_i}{\partial \theta_j} &=& \frac{\partial \binom{N'_i}{h_i}J_s(h_i, \wh_i)\wq_i^{\wh_i}\wtp_i^{s N'_i-\wh_i}}{\partial \theta_j} \nonumber \\
% &=& K_i\left(\frac{\wh_i}{\wq_i}\frac{\partial \wq_i}{\partial \theta_j} + \frac{s N'_i-\wh_i}{\wtp_i}\frac{\partial \wtp_i}{\partial \theta_j}\right) \nonumber \\
% &=& K_i \left(\frac{\wh_i}{\wq_i} - \frac{s N'_i-\wh_i}{\wtp_i}\right) \frac{\partial \wq_i}{\partial \theta_j} \nonumber \\
% &=& K_i \frac{\wh_i - s N'_i \wq_i}{\wtp_i \wq_i} \frac{\partial \wq_i}{\partial \theta_j}.
\end{equation}
So similarly with Eq.~\ref{eq:I1term}, the first term simplifies to
\begin{equation}
\sum_{\whl, \newhl}\wKll\frac{1}{K_l}\frac{\partial K_l}{\partial \theta_i} \frac{\partial K_l}{\partial \theta_j} \nonumber  = s N \frac{F_l}{\wtp_l \wq_l} \frac{\partial \wq_l}{\partial \theta_i}\frac{\partial \wq_l}{\partial \theta_j}.
\end{equation}
The second and third terms can be further simplified, similarly to what was done in the 1-PNR case.
% Eq.\ref{eq:I2term} and Eq.\ref{eq:I3term}, can simplify to
% \begin{eqnarray}
% & & \sum_{\whl, \newhl}\frac{1}{\wKll} K_l \frac{\partial \wKll}{\partial \theta_i} \frac{\partial \wKll}{\partial \theta_j} \nonumber \\
% &=& \sum_{\whll, \hll} P(\whll, \hll;\bm{\theta})\left(\frac{\partial\ln{P(\whll, \hll;\bm{\theta})}}{\partial \theta_i}\right)\left(\frac{\partial\ln{P(\whll, \hll;\bm{\theta})}}{\partial \theta_j}\right)
% \end{eqnarray}
% and
% \begin{equation}
% \sum_{\whl, \newhl}\frac{\partial \wKll}{\partial \theta_i}\frac{\partial K_l}{\partial \theta_j} = 0.
% \end{equation}
Therefore, the final Fisher information matrix is simplified as
\begin{equation}
\label{eq:Is}
I_{ij}(\bm{\theta}) = s N \sum_{k=l_1}^{l_2} \frac{F_k}{\wtp_k \wq_k} \frac{\partial \wq_k}{\partial \theta_i}\frac{\partial \wq_k}{\partial \theta_j}.
\end{equation}
% \begin{equation}
%     P\{\wh(i)=\wh_i|h(i-1)=h_{i-1}, h(i-2)=h_{i-2}, ..., h(i-T)=h_{i-T}\}=\binom{s N'_i}{\wh_i}\wq_i^{\wh_i}\wtp_i^{s N'_i-\wh_i}
% \end{equation}

% \begin{eqnarray}
%     &&E[\wh(i)] \nonumber \\
%     &=& \wq_i \sum_{h_{i-1}=1}^{N}\cdots \sum_{h_{i-T}=1}^{N} s (N-h_{i-1}-\cdots-h_{i-T}) P\{h(i-1)=h_{i-1}, ..., h(i-T)=h_{i-T}\} \nonumber \\
%     &=& s \wq_i (N - E[h(i-1)] - \cdots - E[h(i-T)])
% \end{eqnarray}

% Similarly, we denote $E[\wh(i)] = N s \wq_i F_i$.
Similar to Eq.~\ref{eq:I1}, each term in Eq.~\ref{eq:Is} exactly corresponds to the Fisher information for estimating $\bm{\theta}$ when the count in the $i$-th bin of the histogram follows a binomial distribution $\mathcal{B}(s N F_i(\bm{\theta}), \wq_i(\bm{\theta}))$.
Clearly, when $s = 1$, Eq.~\ref{eq:Is} reduces to the same result as the ranging performance for detectors without photon-number resolution, as given in Eq.~\ref{eq:I1}.

In the above derivation, both the total subpixel-triggered histogram $\bm{\wh}$ and the TDC trigger counts $\bm{h}$ are assumed to be known.
However, many current dToF systems based on multi-subpixel SPAD architectures only record $\bm{\wh}$ without storing $\bm{h}$.
The absence of $\bm{h}$ clearly leads to a loss of Fisher information, which in turn degrades the estimation performance for $t_0$.
To distinguish between these two system architectures, we define systems that record both $\bm{\wh}$ and $\bm{h}$ as Type I, and those that record only $\bm{\wh}$ as Type II.
For Type II systems, the Fisher information becomes
\begin{equation}
    \label{eq:P3}
    I_{ij}(\bm{\theta}) = \sum\limits_{\whl}\left[\sum\limits_{\newhl}P(\whl, \newhl;\bm{\theta})\left(\sum\limits_{\newhl}\frac{\partial\ln{P(\whl, \newhl;\bm{\theta})}}{\partial \theta_i}\right)\left(\sum\limits_{\newhl}\frac{\partial\ln{P(\whl, \newhl;\bm{\theta})}}{\partial \theta_j}\right)\right].
\end{equation}
Since $\bm{h}$ becomes latent variables in this case, the above expression is difficult to simplify into a clear closed form.
Here an asymptotic expression for large $N$ will be derived to evaluate the system performance in the limit.

To derive the asymptotic expression, the random vector $\wh(i)$ is approximated asymptotically by a multivariate normal distribution.
For the expectation, from Eq.~\ref{eq:Ewhi}, $E[\wh(i)] = s N F_i \wq_i$.
Due to the absence of trigger count information $\bm{h}$, whether the $i$-th bin is within the dead time follows a Bernoulli distribution with probability $F_i$.
According to the total variance equation, the variance of $\wh(i)$ is
\begin{eqnarray}
\operatorname{Var}\{\wh(i)\} & = & \operatorname{Var}\{\mathcal{B}(s N F_i,\wq_i)\}+ (s\wq_i)^2 \operatorname{Var}\{\mathcal{B}(N, F_i)\} \nonumber \\
& = & s N F_i \wq_i(1-\wq_i) + s^2 \wq_i^2 N F_i(1-F_i) \nonumber \\
& = & s N C_{ii},
\end{eqnarray}
where $C_{ii} = F_i \wq_i(1-\wq_i) + s \wq_i^2 F_i(1-F_i)$.
For the covariance between $\wh(i)$ and $\wh(i)$, let $Y(i)$ denotes the random variables of SPAD triggered in a single detection event at the $i$-th bin. 
Due to the presence of dead time, when $i\neq j$ and $|i-j|\tau\leq T$, it holds $Y(i) Y(j) = 0$, hence
\begin{eqnarray}
    \operatorname{Cov}(\wh(i), \wh(j)) & = & N \operatorname{Cov}(Y(i), Y(j)) = \mathcal{N}(E[Y(i) Y(j)]-E[Y(i)]E[Y(j)]) \nonumber \\
    & = & -N E[Y(i)]E[Y(j)] = -N s^2 F_i F_j \wq_i \wq_j = s N C_{ij}
\end{eqnarray}
where $C_{ij} = -s F_i F_j \wq_i \wq_j$.
Compared with Type I system, the presence of $F_i(1-F_i)$ term in variance for counts in each bin and covariance between bins in Type II systems prevents direct inference of whether each bin was affected by the dead time during individual measurements based solely on the histogram data $\bm{\wh}$.
This inter-bin correlation reduces the overall obtainable Fisher information.

So in the large-$N$ limit, the joint distribution of $\bm{\wh}=(\wh_1,\ldots,\wh_T) $ converges to a multivariate normal distribution $\mathcal{N}(s N \bm{Q}, s N \bm{C})$ where $\bm{Q} = [Q(0), Q(1), \dots, Q(T)]$.
The Fisher information for the Type II system is ultimately expressed as \cite{kayFundamentalsStatisticalSignal1993}
\begin{equation}
    \label{eq:hFisher}
I_{ij}(\bm{\theta}) = s N \frac{\partial \bm{Q}(\bm{\theta})}{\partial \theta_i}\bm{C}(\bm{\theta})^{-1}\frac{\partial \bm{Q}(\bm{\theta})}{\partial \theta_j} + O(1).
\end{equation}
where the explicit expression of $\bm{C}^{-1}$ is shown in Appendix B.
Since the first term is proportional to $s N$, it dominates when $N$ is large.

In conclution, the above derivations show that the CRLB for estimating $t_0$, $\operatorname{Var}\{\widehat{t_0}\}$, depends on both $t_0$ and $R$ under all three system architectures: 1-PNR SPAD and Type I/II $s$-PNR SPAD systems.
In practical measurements, although the surface reflectivity of the target cannot be controlled, the system can adjust $R$ by varying the laser output power, thereby tuning the estimation performance of $t_0$.
To validate the theoretical results, Monte Carlo simulations were conducted to generate histograms for different system architectures, followed by parameter estimation using the maximum likelihood method [see Appendix C].
Figure~\ref{fig:typical-vart0} presents typical CRLB behaviors for estimating $t_0$ under varying $R$ for three detection architectures.
Since the CRLB scales inversely with the number of detection events $N$ for all three architectures, to facilitate intuitive comparison of ranging performance at different $N$, we define $\delta t_0 = \sqrt{N\times\operatorname{Var}\{\widehat{t_0}\}} = 1/\sqrt{\mathcal{I}_{11}(1-\rho^2)}$, representing the standard deviation of the ToF measurement per single pulse measurement at given $t_0$ and $R$.
It can be seen from the figure that the results show excellent agreement between Monte Carlo simulations combined with maximum likelihood estimation and the theoretical CRLB predictions, confirming the validity of the theoretical analysis.

\begin{figure}[!h]
\centering % Only used for preprints
\includegraphics[width=7 cm]{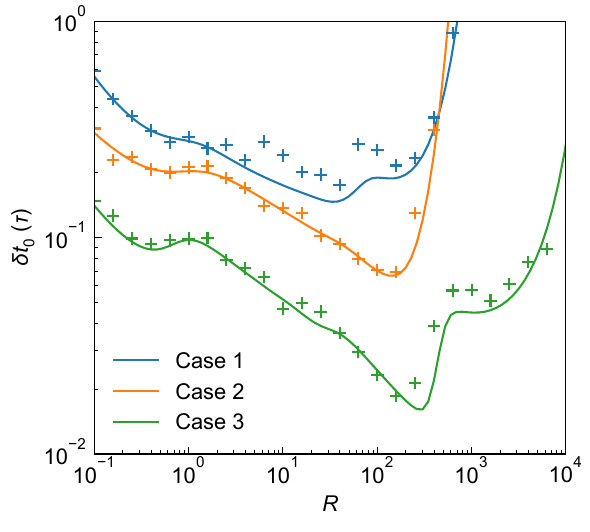}
\caption{Theoretical ranging performance $\delta t_0$ normalized to TDC resolution $\tau$ for estimating the ToF $t_0$ as a function of the peak received photon flux rate $R$ under three system architectures.
Solid lines represent theoretical calculations, and crosses denote the variances of 100 sets of $t_0$ estimates obtained via maximum likelihood estimation from histograms generated by Monte Carlo simulations at different values of $R$.
$f(t)$ is assumed as a gaussian-shaped laser pulse with full width at half maximum (FWHM) $w=4\tau$.
Case 1 (blue): 1-PNR SPAD with $N=100$, $t_0 = 10\tau$ and $R_n=0.02$;
Case 2 (range): 4-PNR SPAD with Type I system, $N=100$, $t_0 = 10.2\tau$ and $R_n = 0$;
Case 3 (green): 4-PNR SPAD with Type II system, $N=1000$, $t_0 = 10.5\tau$ and $R_n = 0.01$.
\label{fig:typical-vart0}}
\end{figure}   
\unskip

%%%%%%%%%%%%%%%%%%%%%%%%%%%%%%%%%%%%%%%%%%
\section{Results}

The following section utilizes the analytic CRLB from the previous chapter to further analyze the ranging performance of different system architectures and investigate how the systems can achieve their fundamental ranging limits.

\subsection{1-PNR SPAD}

\subsubsection{Performance Evaluation at a Fixed Range}
\begin{figure}[!h]
\centering % Only used for preprints
\includegraphics[width=7 cm]{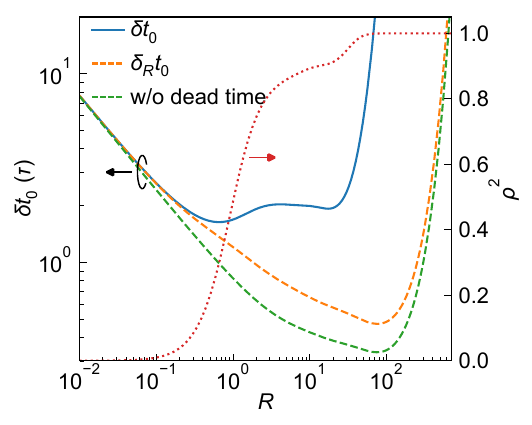}
\caption{System ranging performance for a gaussian-shaped laser pulse with FWHM $w=3\tau$ and ToF with $t_0 = 10\tau$ as a function of $R$.
Variation of system ranging precision $\delta t_0$ (blue solid, left axis), ranging precision with known $R$, $\delta_R t_0$ (orange dashed, left axis), ranging precision without accounting for dead time (green dashed, left axis), and correlation coefficient $\rho^2$ (red dotted line, right axis) are shown as functions of $R$ for comparison.
\label{fig:vart0-vs-R}}
\end{figure}

Figure~\ref{fig:vart0-vs-R} illustrates the dependence of the system ranging precision $\delta t_0$ on the peak received photon flux rate $R$, under the condition where the gaussian-shaped laser pulse has a FWHM of $3\tau$ and the true photon arrival time $t_0=10\tau$.
For comparison, the figure also plots the lower bound $\delta_R t_0 = 1/\sqrt{\mathcal{I}_{11}}$, which represents the standard deviation limit for estimating $t_0$ assuming a known $R$, as well as the performance neglected the effect of detector dead time (i.e., let $F_i=1$ for all bins).
It can be observed that when the effect of dead time is neglected, the ranging performance curve lies at the bottom, achieving a minimum value of $\delta t_0 \approx 0.33\tau$, which is approximately five times better than the best performance when dead time is considered in this particular case ($\delta t_0 \approx 1.63\tau$), indicating an overly optimistic estimation of the system's ranging capability.
This discrepancy also highlights the adverse impact of walk error on achievable ToF accuracy.

Specifically, in the regime of small $R$, the ranging precision predicted by all three conditions improves with increasing $R$, which is reasonable:
at low photon flux rate, the pixel is rarely triggered, resulting in very limited information content, and a modest increase in $R$ significantly enhances the likelihood of photon detection, thereby improving the Fisher information for estimating $t_0$.
Moreover, due to the negligible effect of temporal pile-up with small $R$, the correction factor $F_i\approx1$ and $\rho\approx0$, leading to nearly identical ranging precision curves predicted by all three conditions.

In contrast, as $R$ becomes very large, $\delta_R t_0$ diverges, indicating that even with known $R$, accurate estimation of $t_0$ becomes fundamentally infeasible.
Intuitively, when $R$ is extremely high, nearly all detection events trigger in the first histogram bin where the laser pulse falls, making the peak effectively only one bin wide.
This extreme pile-up effect limits the amount of available temporal information, preventing sub-bin estimation of $t_0$.

Notably, although $\delta_R t_0$ may still appear to decrease when $R > 0.5$, $\delta t_0$ begins to plateau and eventually diverges beyond $R \approx 30$.
This behavior is further explained by the evolution of $\rho^2$, which begins to increase rapidly beyond $R \approx 0.1$ and asymptotically approaches unity.
The increase in $\rho^2$ indicates the emergence and strengthening of pile-up effects, leading to a growing statistical correlation between $t_0$ and $R$, which in turn degrades the precision of $t_0$ estimation and causes $\delta t_0$ to diverge.

In the intermediate region, approximately $0.1 \lesssim R \lesssim 40$, the variation of $\delta t_0$ arises from a delicate interplay between $\delta_R t_0$ and the correlation factor $\rho^2$, resulting in a non-monotonic and complex dependence.
As shown in the figure, $\delta t_0$ exhibits two local minima, with the leftmost minimum occurring near $R \approx 0.64$ and reaching a value of $\delta t_0 \approx 1.63\tau$.
This suggests that, to approach the theoretical limit of ranging performance under this specific condition, the laser output power should be carefully tuned such that the received photon flux rate $R$ lies near this optimal value.

\unskip

\begin{figure}[!h]
\centering % Only used for preprints
\includegraphics[width=14 cm]{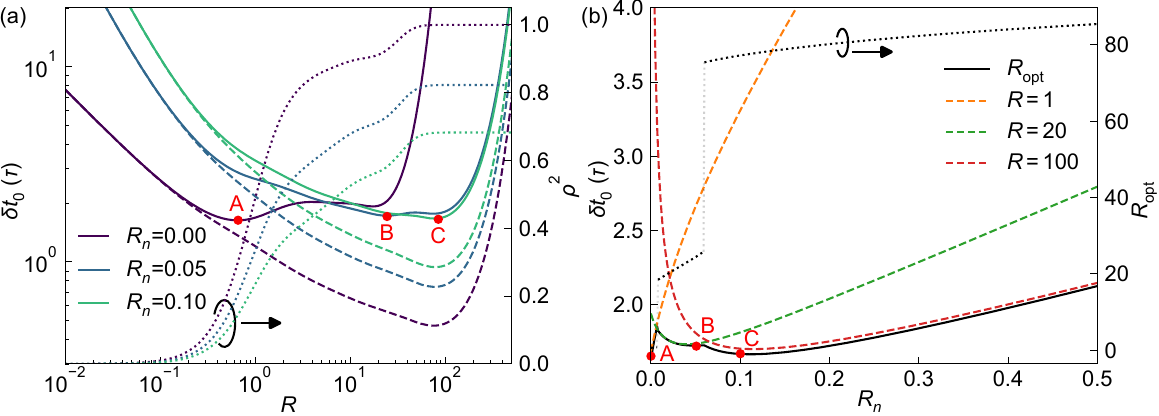}
\caption{System ranging performance as a function of background noise intensity $R_n$ with $w=3\tau$ and $t_0 = 10\tau$.
(a) Variation of system ranging precision $\delta t_0$ (solid line, left axis), ranging precision with known $R$, $\delta_R t_0$ (dashed line, left axis), and correlation coefficient $\rho^2$ (dotted line, right axis) as functions of $R$, under different $R_n$ levels.
(b) $\delta t_0$ as a function of $R_n$ for representative values of $R$.
Black solid: optimal ranging precision under each $R_n$; black dotted: corresponding optimal $R=R_\mathrm{opt}$;
Orange dashed: $R=1$;
Green dashed: $R=20$;
Red dashed: $R=100$;
Points A, B, and C in both subfigures mark the optimal ranging performance points for different $R_n$ values.
\label{fig:vart0-vs-Rbg}}
\end{figure}

Figure~\ref{fig:vart0-vs-Rbg}(a) further illustrates the variation of ranging precision $\delta t_0$ as a function of $R$ under different background noise flux rate $R_n$, with ToF $t_0=10\tau$ fixed.
It can be observed that as $R_n$ increases, the lower bound $\delta_R t_0$ (dashed lines) generally increases, indicating that the information available for estimating $t_0$ alone reduces due to the increased background noise.
On the other hand, with increasing $R_n$, the correlation coefficient $\rho$ decreases overall, implying a mitigation of pile-up effects and thus a weakened statistical correlation between $t_0$ and $R$.
Intuitively, the pile-up effect of the laser pulse manifests as a high count rate at the rising edge, causing the latter half of the pulse to almost entirely fall within the detector dead time.
The presence of background noise introduces a nonzero probability that the SPAD is already in dead time triggered by background counts when the pulse arrives, thereby allowing the detector to recover just in time to respond to the latter half of the pulse, alleviating the pile-up effect.
As a result of the combined influence of $\delta_R t_0$ and $\rho$, the system's ranging precision $\delta t_0$ at lower $R$ deteriorates with increasing $R_n$, whereas at higher $R$ it actually improves.
Figure~\ref{fig:vart0-vs-Rbg}(b) shows the dependence of $\delta t_0$ on $R_n$ at two fixed detection rates with $R=1$ and $R=100$, more clearly demonstrating this trend.

Figure~\ref{fig:vart0-vs-Rbg}(b) also depicts the system's optimal ranging precision and the corresponding optimal received photon flux rate $R_\opt$ as functions of $R_n$.
It can be seen that while the optimal ranging precision remains approximately constant with increasing $R_n$ (around 1.6 to 2.1), the optimal detection rate $R_\opt$ undergoes two abrupt jumps near $R_n \approx 0.01$ and $0.06$, increasing first from about 1 to 20 and then from 20 to roughly 80, indicating a significant increase in the required photon flux.
These jumps occur because the CRLB curve develops new minima around $R \approx 20$ at $R_n \approx 0.01$ and $R\approx 100$ at $R_n \approx 0.06$, as a result of delicate interplay between $\delta_R t_0$ and $\rho^2$ [see points A, B, and C in Fig.~\ref{fig:vart0-vs-Rbg}(a)].

To avoid excessive echo power requirements that could strain the system, one may consider the ranging performance constrained by the photon flux upper bound.
For example, Fig.~\ref{fig:vart0-vs-Rbg}(b) presents $\delta t_0$ constrained by $R<20$, corresponding to the optimal detection rate after the first jump.
Although the system's ranging performance degrades beyond the theoretical optimum for $R_n > 0.06$, the maximum deterioration is about 30\%, which can serve as a practical reference for the trade-off between photon flux and ranging capability in system design.

\unskip

\subsubsection{System Optimization for Optimal Ranging Precision}
In the absence of prior knowledge about the target distance, it may seem necessary to evaluate the ranging precision over all possible values of $t_0$ to assess overall system performance and enable further optimization.
However, the following analysis demonstrates that it suffices to consider only the case where the received laser pulse lies within the first bin, i.e., $t_0 \in [0, \tau)$, to generalize the ranging performance over the entire histogram range.

From Eq.~\ref{eq:IijFl}, it is clear that for a given system and fixed $R$, the variance of the ToF estimate $\delta t_0$ is determined solely by the values of $F_i$ associated with the bins occupied by the received laser pulse.
Moreover, Eq.~\ref{eq:F0def} indicate that the values of $F_i$ are determined by the value of $F_{l_1}$ corresponding to the first bin occupied by the received laser pulse, the photon detection probabilities $q_i$ across the laser pulse, and the background photon triggering probability $q_b$.
For systems employing a single-TDC architecture, when $t_0$ is shifted by an integer number of bins, the ratio $F_i / F_{l_1} = \prod_{j=l_1}^{i-1} p_j$ remains invariant.
According to Eq.~\ref{eq:vart0}, since $\rho^2$ is homogeneous with respect to $F_i$, it remains unchanged as well, and therefore $\delta t_0 {|}_{t_0 = (l_1 + \varepsilon) \tau} = \sqrt{F_0/F_{l_1}}\cdot \delta t_0{|}_{t_0 = \varepsilon \tau}$.
Equation~\ref{eq:F0def} implies $F_{l_1 > T} = F_0(1 - q_b)^{l_1 - T}$, and thus the ranging precision for the single-TDC architecture decreases exponentially with increasing leading bin inde $l_1$ under a fixed photon flux rate $R$, expressed as
\begin{equation}
    \delta t_0 {|}_{t_0 = (l_1 + \varepsilon) \tau} = (1-q_b)^{l_1-T}\delta t_0{|}_{t_0 = \varepsilon \tau}.
\end{equation}
For a multi-event TDC architecture, Eq.~\ref{eq:Frec} implies that $F_{i \leq l_1} = 1/(1+q_b T)$, i.e., all $F$ values prior to the arrival of the laser pulse are identical, thereby leading to the relation
\begin{equation}
    \delta t_0 {|}_{t_0 = (l_1 + \varepsilon) \tau} = \delta t_0{|}_{t_0 = \varepsilon \tau}.
\end{equation}
which means under a fixed $R$, the ranging precision remains unchanged when $t_0$ is shifted by an integer number of bins.
Therefore, the overall ranging capability of the system across the full ToF range can be fully characterized by analyzing $\delta t_0$ within the single-bin interval for both single- and multi-event TDC architectures.

\begin{figure}[!h]
\centering
\includegraphics[width=7 cm]{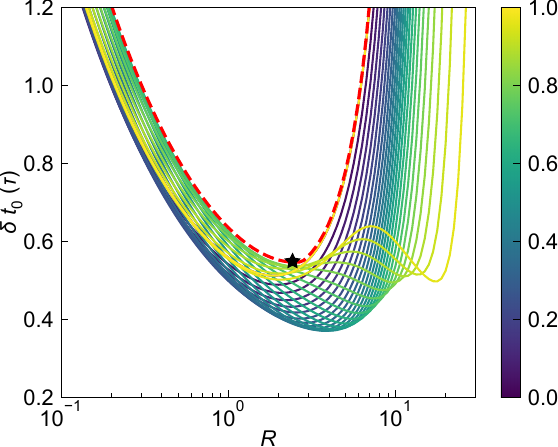}
\caption{
    Ranging precision $\delta t_0$ as a function of $R$ for different values of $t_0$ varying from 0 (purple) to $\tau$ (yellow) with laser pulse width of $0.6\tau$.
    The worst-case ranging precision over $t_0 \in [0, \tau)$, denoted as $\Delta t_0$, is shown as a red dashed line.
    The minimum standard deviation $\Delta_\mathrm{m} t_0$ and the corresponding optimal received photon flux rate $R_\opt$ are marked by a black star.
\label{fig:vart0-vs-R-vs-t0}}
\end{figure}

Figure \ref{fig:vart0-vs-R-vs-t0} shows how $\delta t_0$ varies with $R$ for different values of $t_0$ within a single bin $[0, \tau)$.
It is evident that while the absolute values differ slightly, the overall trend remains consistent:
$\delta t_0$ decreases initially with increasing $R$, then increases again.
Notably, for $t_0 = 0$ and $t_0 = \tau$, the curves coincide exactly due to the symmetry introduced by a full-bin shift.
Since no prior knowledge of the target ToF is available, i.e., $t_0$ can take any value within $[0,\tau)$, to ensure the system's ranging performance under the worst-case scenario, the system's ranging performance at a given $R$ is defined as the maximum value of $\delta t_0$ over $t_0\in[0,\tau)$, denoted as
\begin{equation}
    \Delta t_0 = \max_{\varepsilon\in[0,1)}\{{\delta t_0} {|}_{t_0 = \varepsilon \tau}\},
\end{equation}
corresponding to the upper envelope in Fig. \ref{fig:vart0-vs-R-vs-t0} (red dashed line).
It can be observed that $\Delta t_0$ at different values of $R$ corresponds to the $\delta t_0$ over different $t_0$, making it difficult to derive a closed-form expression, and thus must be obtained numerically.
In practical systems, by tuning the laser power to adjust $R$, the system can be operated at the minimum of $\Delta t_0$, denoted as $\Delta_\mathrm{m} t_0$, and the corresponding optimal photon arrival rate $R_\opt$, as marked by the black star in the figure.
This point can be regarded as the fundamental ranging precision limit of the system under complete lack of prior knowledge about the target.

\unskip

\begin{figure}[!h]
\centering % Only used for preprints
\includegraphics[width=14 cm]{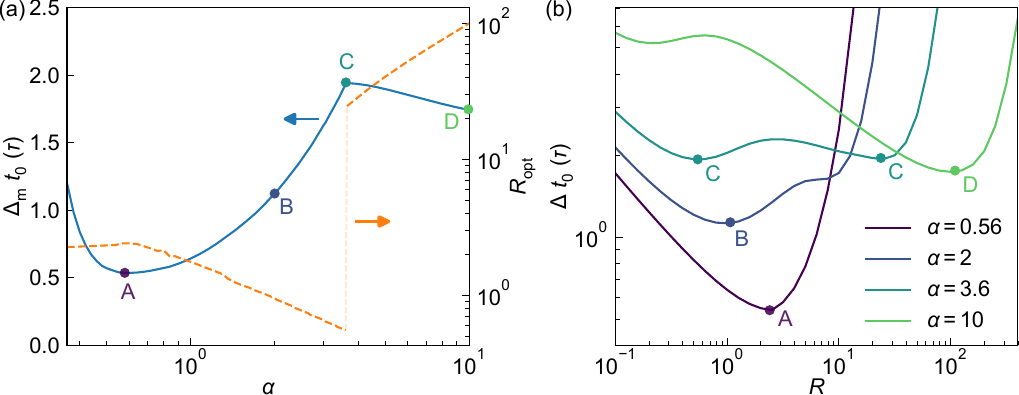}
\caption{Ranging performance of the system as a function of laser pulse width $w=\alpha\tau$.
(a) Theoretical optimal ranging precision $\Delta_\mathrm{m} t_0$ (solid blue) and the corresponding optimal detection rate $R_\opt$ (dashed orange) for different $\alpha$.
The translucent section of the $R_\opt$ curve indicates the region of abrupt transition.
Points A-D correspond to the minima under $\alpha = 0.56$, 2, 3.6, and 10, respectively.
(b) Ranging precision $\Delta t_0$ as a function of $R$ for the four representative pulse widths in (a), along with the respective optimal points A-D.
\label{fig:vart0-vs-s}}
\end{figure}

One of the key design parameters in a dToF system is the ratio between the laser pulse width $w$ and the TDC resolution $\tau$.
Figure~\ref{fig:vart0-vs-s}(a) illustrates the system's optimal ranging performance $\Delta_\mathrm{m} t_0$ (normalized to $\tau$) as a function of laser pulse width $w = \alpha\tau$ under a fixed $\tau$.
As $\alpha$ approaches zero, the laser pulse energy becomes almost entirely concentrated within a single bin, and the histogram fails to capture sufficient information to support sub-bin precision, which results in the divergence of the ranging variance.
As $\alpha$ increases, more timing information becomes available from the histogram, and the ranging precision improves.
The optimal performance is achieved at $\alpha \approx 0.56$, where $\Delta_\mathrm{m} t_0$ reaches a minimum of approximately $0.53\tau$ at $R_\opt \approx 2.4$.
In practical terms, for a TDC resolution of 1~ns, a single pulse measurement under a laser pulse with width of 560~ps corresponds to a rannging standard deviation of about 300~ps, or roughly 45~mm, theoretically, and can be futher improved to 4.5~mm on the accumulated histogram from 100 pulses.
As $\alpha$ increases beyond this point, $\Delta_\mathrm{m} t_0$ gradually worsens. 
However, as shown in Fig.~\ref{fig:vart0-vs-s}(b), a secondary minimum emerges at higher $R$ (at $R\approx 7$ in curve B), and this second minimum decreases as $\alpha$ increases.
When $\alpha \approx 3.64$, both minima converge to approximately 1.90.
With further increases in $\alpha$, the secondary minimum becomes the global minimum, yielding improved system performance again. Nevertheless, the corresponding optimal $R_\opt$ jumps from around 2 to approximately 30, implying a more than tenfold increase in required laser power, which may impose significantly higher demands on the laser emission capability of the dToF system.
Therefore, in practical dToF system design, the pulse width should be kept with an ideal value near $0.56\tau$ to achieve the best overall ranging precision.

\unskip

\begin{figure}[!h]
\centering % Only used for preprints
\includegraphics[width=14 cm]{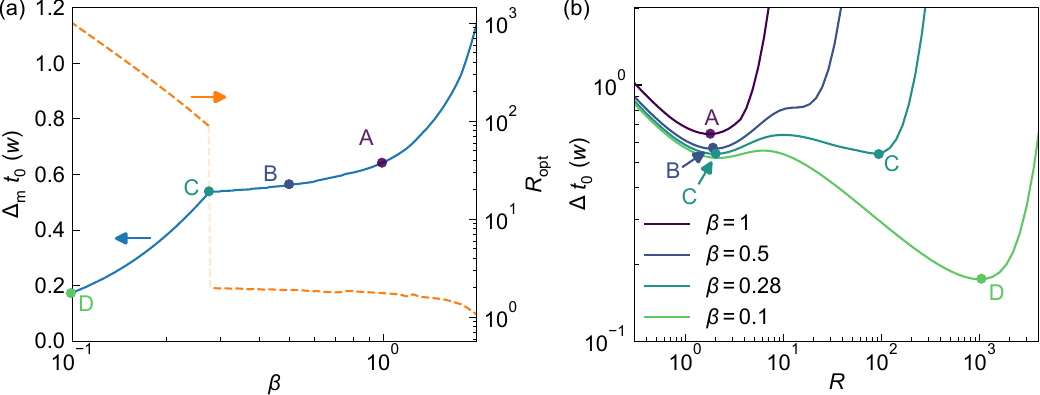}
\caption{System ranging performance as a function of TDC resolution $\tau = \beta w$.
(a) Theoretical optimal ranging precision $\Delta_\mathrm{m} t_0$ (solid blue) and the corresponding optimal photon rate $R_\opt$ (dashed orange) under varying $\beta$.
The translucent section of the $R_\opt$ curve indicates the region of abrupt transition.
Points A-D correspond to the minima under $\beta = 1$, 0.5, 0.28 and 0.1, respectively.
(b) $\Delta t_0$ as a function of $R$ at representative values of $\beta$, along with their optimal performance points A-D.
\label{fig:vart0-vs-tau}}
\end{figure}

When the laser pulse width $w$ is fixed, Fig.~\ref{fig:vart0-vs-tau}(a) illustrates the system's ranging performance $\Delta_\mathrm{m} t_0$ (normalized to $w$) and the corresponding optimal detection rate $R_\opt$ under different TDC resolutions $\tau = \beta w$.
For large $\beta$, the histogram bin width exceeds the laser pulse width, making it difficult to extract sufficient timing information from the histogram for sub-bin accuracy.
As a result, $\Delta_\mathrm{m} t_0$ tends toward infinity.
As $\beta$ decreases, temporal resolution of the pulse improves, leading to a steady decrease in $\Delta_\mathrm{m} t_0$.
During this regime, although $R_\opt$ increases, it remains within a moderate range (typically between 1 and 2).
Moreover, when $\beta < 1$, the curve of $\Delta_\mathrm{m} t_0$ begins to flatten, indicating diminishing returns from further improvements in TDC resolution.
For example, improving TDC resolution from $\beta = 0.6$ to $\beta = 0.3$ (i.e., halving $\tau$) reduces $\Delta_\mathrm{m} t_0$ from $0.57w$ to about $0.54w$, an improvement of only \textasciitilde 5\%.

On the other hand, Fig.~\ref{fig:vart0-vs-tau}(b) shows that at larger $R$ values ($R_\opt > 100$), an alternative local minimum in $\Delta t_0$ emerges and continues to decrease with smaller $\beta$.
This behavior resembles that observed in Fig.~\ref{fig:vart0-vs-s}, which is reasonable since increasing $\tau$ with fixed $w$ is equivalent to decreasing $w$ with fixed $\tau$.
At around $\beta \approx 0.28$, the larger-$R$ local minimum becomes the global minimum and continues to drop rapidly as $\beta$ decreases.
As $\beta$ approaches 0, the system effectively samples the return pulse in a nearly continuous manner, allowing $\Delta_\mathrm{m} t_0$ to approach zero.
However, since the bin width $\tau$ also tends toward zero in this limit, achieving meaningful photon counts per bin would require an infinite photon flux rate, which is reflected in the divergence of $R_\opt$, and is clearly impractical for real-world applications.

Therefore, when the laser pulse width $w$ is fixed, a TDC resolution slightly smaller than $w$ is sufficient to achieve near-optimal system performance without requiring excessively high $R$.
Further reducing $\tau$ below $0.3w$ may yield marginal theoretical gains in precision, but these come at the cost of rapidly increasing photon demands and greater susceptibility to practical limitations such as TDC nonlinearity \cite{xiaReviewAdvancementsTrends2024} and timing jitter \cite{liuEfficientComprehensiveTiming2024,xuEfficientSimplifiedSPAD2025}, making the theoretical optimum difficult to attain in practice.

\unskip

\subsection{SPAD detector with photon-number resolution}

\begin{figure}[!h]
    \centering % Only used for preprints
    \includegraphics[width=7 cm]{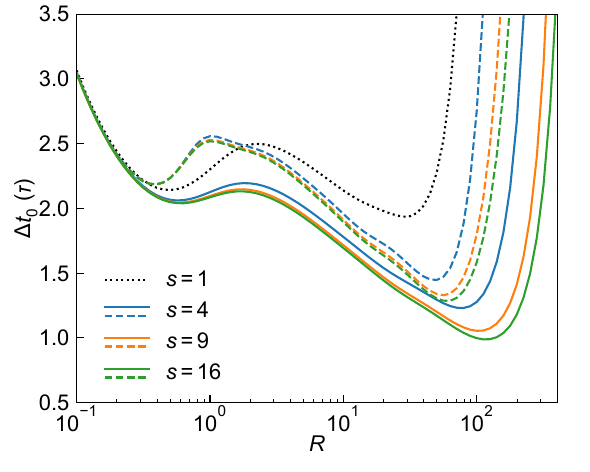}
    \caption{Ranging precision $\delta t_0$ versus recived photon flux rate $R$ for 1-PNR SPAD (black dotted), 4-PNR (blue), 9-PNR (orange) and 16-PNR (green) detectors under Type I (solid) and Type II (dashed) systems.
    \label{fig:vart0-vs-sh}}
\end{figure}

Figure~\ref{fig:vart0-vs-sh} illustrates the theoretical optimal ranging precision $\Delta_\mathrm{m} t_0$ as a function of the photon rate $R$, using 4-PNR (e.g., $2 \times 2$ subpixels), 9-PNR (e.g., $3 \times 3$), and 16-PNR SPAD (e.g. $4\times 4$) macro-pixels under both Type I and Type II system architectures, with the laser pulse width set to $w = 2\tau$.
Compared to the 1-PNR case, increasing the number of subpixels, i.e., improving photon-number resolution, shifts the $\Delta_\mathrm{m} t_0$ curves downward and to the right, indicating improved theoretical performance also a higher required return photon rate $R$.
Moreover, as the number of subpixels $s$ increases from 1 to 16, the overall shift of the curves gradually diminishes, indicating diminishing returns in performance improvement.
In addition, for a fixed $s$, Type II system consistently yields worse $\Delta_\mathrm{m} t_0$ than Type I, which is is consistent with our previous analysis in Eq.~\ref{eq:hFisher}:
the lack of TDC trigger count information in Type II leads to stronger inter-bin correlations due to dead time, increasing the covariance and reducing the overall Fisher information.

\unskip

\begin{figure}[!h]
    \centering % Only used for preprints
    \includegraphics[width=14 cm]{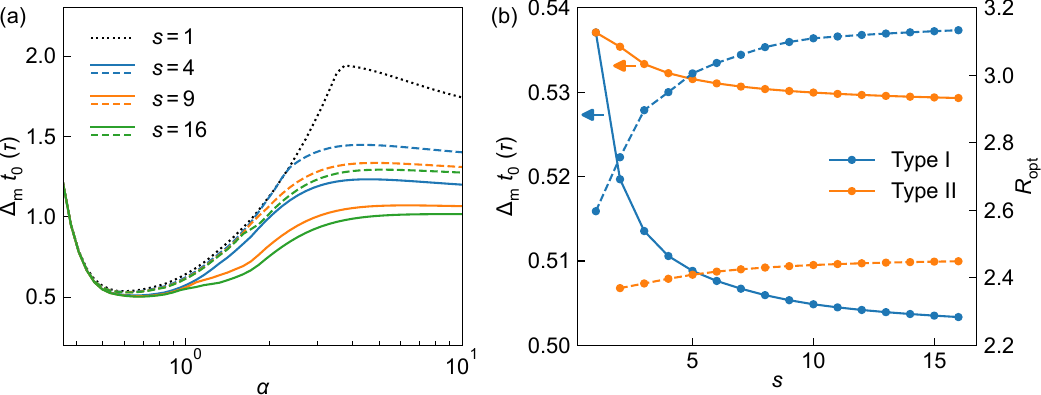}
    \caption{System ranging performance as a function of the number of subpixels $s$.
    (a) Theoretical optimal ranging precision $\Delta_\mathrm{m} t_0$ versus normalized laser pulse width $\alpha$ for 1- (black dotted), 4- (blue), 9- (orange), and 16-PNR (green) detectors under Type I (solid) and Type II (dashed) architectures.
    (b) $\Delta_\mathrm{m} t_0$ (solid) and the corresponding optimal photon rate $R_\opt$ (dashed) as functions of subpixel number $s$ for Type I (blue) and Type II (orange) architectures.
    \label{fig:sh-sat}}
\end{figure}

Figure~\ref{fig:sh-sat}(a) further shows the theoretical optimal ranging precision $\Delta_\mathrm{m} t_0$ and the corresponding optimal photon flux rate $R_\opt$ as functions of the normalized pulse width $\alpha = w/\tau$ for various values of $s$, under Type I and II systems.
When $\alpha$ is large, the required $R_\opt$ is also high, and increasing $s$ leads to a clear improvement in ranging precision, with Type I consistently outperforming Type II.
In contrast, for $\alpha < 1$, $R_\opt$ remains low, and the benefit of higher photon-number resolution becomes marginal, as evidenced by the nearly overlapping $\Delta_\mathrm{m} t_0$ curves across different values of $s$.
Figure~\ref{fig:sh-sat}(b) shows the minimal $\Delta_\mathrm{m} t_0$ achieved at the optimal $\alpha$ for each $s$, which reflects the improvement of the theoretical ranging performance limit enabled by photon-number-resolving capability.
For Type I configuration, increasing $s$ from 1 to 16 improves $\Delta_\mathrm{m} t_0$ from approximately $0.536\tau$ to $0.503\tau$ — an improvement of less than 10\%.
The marginal benefit quickly decreases: increasing $s$ from 1 to 9 yields a $\sim9.4$\% improvement, but going from 9 to 16 only adds $\sim0.5$\%.
For Type II configuration, the gains are even smaller: increasing $s$ from 1 to 16 improves $\Delta_\mathrm{m} t_0$ by only $\sim1.3$\%, from $\sim0.536\tau$ to $\sim0.529\tau$.
The same figure also plots the optimal photon rate $R_\opt$ versus $s$.
Although $R_\opt$ does increase with $s$, it remains in the modest range of $2\sim3$, indicating that the high dynamic range potential of large-$s$ PNR SPADs is not fully utilized, which further limits the performance benefits brought by additional subpixels.

\unskip

In summary, while SPAD detectors with photon-number resolution can improve system dynamic range and maintain good performance under high photon flux rate, the theoretical optimal ranging precision is typically achieved under relatively low pulse widths and moderate photon flux rates.
As a result, the marginal benefit of increasing subpixel count diminishes rapidly.
Moreover, practical issues such as inter-subpixel mismatch and optical/electrical crosstalk across subpixels may degrade overall macro-pixel performance as $s$ increases \cite{incoronatoStatisticalModellingSPADs2021a}, highlighting the importance of carefully balancing design trade-offs in system implementation.

%%%%%%%%%%%%%%%%%%%%%%%%%%%%%%%%%%%%%%%%%%
\section{Conclusions}

In conclusion, this work theoretically derives the Cramér-Rao lower bound for the ranging performance of dToF LiDAR systems considering the effect of dead time, and further extends the analysis to SPAD detectors with photon-number-resolving capabilities.
The theoretical results agree well with Monte Carlo simulations, validating the correctness of the derivation.
According to the analysis, the pile-up effect caused by dead time not only reduces the information content of each histogram bin but also introduces coupling between the time-of-flight and the received photon flux rate, further degrading the ranging performance.
Because of this coupling, the received photon flux rate exhibits an optimal value rather than the traditionally assumed monotonic performance improvement with increasing photon flux;
moreover, ambient background light can, under certain conditions, enhance system performance.
Further analysis is conducted to determine the optimal laser pulse width and TDC resolution for achieving the theoretical ranging limit.
For SPAD detectors with subpixels, the theoretical analysis shows that dToF systems should record TDC trigger events to fully exploit the information gain provided by photon-number-resolving capability, thereby achieving the theoretical optimal ranging performance.
However, while photon-number resolution can improve the system's dynamic range, its impact on the theoretical limit of ranging precision is quite limited.
We believe the theoretical results derived here represent the fundamental lower bound of ranging performance for dToF systems and can serve as valuable guidance for system design and optimal operating point selection.

\bibliographystyle{unsrt}
\bibliography{ref.bib}

\end{document}